# Strong charge ordering above room temperature in B-site disordered electron-doped manganite SrMn$_{0.875}$Mo$_{0.125}$O$_{3-\delta}$

Aga Shahee* and N. P. Lalla

UGC-DAE Consortium for Scientific Research, University campus, Khandwa road
Indore, India- 452017

## Abstract

Low as well as high-temperature electron and x-ray diffraction studies have been carried out on a rare-earth free B-site disordered electron-doped manganite SrMn$_{0.875}$Mo$_{0.125}$O$_{3-\delta}$ in the temperature range of 83K to 637K. These studies reveal the occurrence of strong charge ordering (CO) at room temperature in a pseudo tetragonally distorted perovskite phase with space-group *Pmmm*. Non integral modulation vector of 8.95 times along [-110] indicates a charge density wave type modulation. The CO phase with basic perovskite structure *Pmmm* transforms to a charge disorder cubic phase through a first order phase transition at 355K. Supporting temperature dependent measurements of resistance and magnetization show a metal-insulator and antiferromagnetic transitions across 355*K* with a wide hysterisis ranging from 150*K* to 365*K*. The occurrence of pseudo tetragonality of the basic perovskite lattice with *c/a* < 1 together with charge-ordered regions with 2-dimensional modulation have been analyzed as the coexistence of two CO phases with $3d_x^2/3d_y^2$ type and $3d_{x^2-y^2}^2$ type orbital ordering.

**Introduction:**

The study of charge-ordering (CO) is of prime importance in the domain of transition metal oxides because it relates with the physics of two important phenomena, the high-$T_c$ superconductivity in cuperate[1,2] and the colossal magnetoresistance (CMR) in hole doped rare-earth based multivalent manganites $Ln_{1-x}A_xMnO_3$ (Ln= lanthanide, A- Alkaline rare)[3,4]. Charge, orbital, spin and lattice degrees of freedom invariably result in rich but complex phase diagram of manganites[5-8]. The role of charge-ordering in stabilizing antiferromagnetic (AFM) ground state is the key to the CMR property in manganites[9-12]. It is the field induced percolative melting of this AFM insulating CO phase into ferromagnetic (FM) metallic phase, which gives rise to CMR effect[3]. Initially CO was observed in half doped manganites, then in other compositions also[13-27]. The occurrence of charge ordering has been reported to be both, commensurately or incommensurately modulated[13-27]. Earlier CO was broadly considered as decoration of $Mn^{+3}$ and $Mn^{+4}$ ions in strip form[13,14], but following the subsequent developments[28-31] the strict ionic picture of CO phase has now evolved as generalized picture of a charge density wave (CDW) of $e_g$-electrons. The CO modulation vector q is found to be closely related to the composition ($x$), $q \approx 1-x$. The crystallographic superstructure of charge and orbital ordered phases has been investigated by different groups using different diffraction techniques but the structure of CO phase largely remains unsettled[14, 16,17,21].

Most of the manganites (e.g. $Pr_{1-x}Ca_xMnO_3$ & $La_{1-x}Ca_xMnO_3$)[13-27], the charge-orbital ordered (COO) phase is observed below room temperature (i.e. $T_{CO} < RT$) except for Bi based manganites $Bi_{1-x}Sr_xMnO_3$[32] and $Bi_{1-x}Ca_xMnO_3$[33,34]. Keeping in view the importance of room temperature CMR manganites, CO transition above room temperature always deserve special attention. The stabilization of CO above room temperature in Bi based manganites is correlated with partial hybridization of the $6s^2$ lone pair of $Bi^{+3}$ ion with $O^{-2}$ p-orbitals, thus lowering the mobility of $Mn^{+3}$ $e_g$ electrons and favoring charge localization and ordering[35]. It should be noted that a suitable B-site substitutional disorder may directly enhance the scattering of $e_g$ electrons, lowering its mobility and resulting in localization[36]. Although the structural, magnetic and transport properties of CO phases in A-site substituted rare earth based solid solution manganites have been extensively studied[37,38] but very little is known about rare earth free and/or B-site substituted electron doped regime of manganites like $SrMnO_3$[39,40]. B-site substitutions have been done in half doped manganites but they are invariably known to stabilize ferromagnetic metallic phase[41-45] and suppress CO. But contrary to that Martin et. al[46] observed stabilization of room temperature CO phase in electron doped and pure alkaline earth based manganite $CaMn_{1-x}M_xO_3$ by substitution of high oxidation state ions M = $Mo^{6+}$ and $Ru^{5+}$ (for $x$=0.07 and 0.12). Due to the high valence of $Mo^{6+}$ ion large variation in Mn valence is possible while keeping its effect on structure minimal[46]. Experimental observation of charge-orbital ordering phenomenon in a B-site disordered electron doped manganite by a $d^0$ ions, like $Mo^{6+}$ substituted at $Mn^{4+}$ site, gives new insight regarding the understanding of the phenomenon of COO. But such reports are sparse, infact none after ref.46. Keeping this in view more such studies are needed. In this regard $SrMn_{1-x}Mo_xO_3$ for $x$<0.2 has been studied[47,48,49] for electrical transport and structural phase transformation behavior. It does show occurrence of metal-insulator and orbital-ordering but they were finally concluded to be charge disordered[47]. In the following we report detailed structural phase-transition studies on $SrMn_{0.875}Mo_{0.125}O_{3-\delta}$ and show the evidence of strong charge-ordering with $T_{CO}$ =355K and bring out some very important differences related to the type of orbital ordering in this.

**Experimental:**

$SrMn_{0.875}Mo_{0.125}O_{3-\delta}$ was prepared through conventional solid-state reaction rout using 99.99% pure $SrCO_3$, $MnO_2$ and $MoO_3$. The stoichiometric mixture of the ingredients was thoroughly mixed using mortar pastel for effectively ~24 hrs and then calcined at $1100^oC$ for 24hrs. The calcined powder was reground and fired at $1400^oC$ for 24hrs. The fired powder was reground and pelletized in the form of ~14mm x 1mm disks and sintered at $1400^oC$ for 48hrs. After sintering the furnace was cooled with a rate of $2^oC/min$. The as prepared sample was then subjected to phase-purity, compositional homogeneity and oxygen stoichiometry characterizations employing powder x-ray diffraction (XRD), EDAX analysis in transmission electron microscope (TEM) and iodometric titration respectively. The XRD was carried out using a Rigaku diffractometer (D-max) equipped with a graphite (002) monochromator. Its angular resolution ($\Delta\theta$) was $0.047^o$ ($8.33 \times 10^{-4}$ rads.) and it was mounted on a rotating anode source producing Cu $K_\alpha$ x-rays at 11Kw. The diffractometer was equipped with $LN_2$ based cryostat and Pt-heater based high-temperature attachments. For composition analysis EDAX was carried out at various close by (~50nm) points in a single grain using a ~20nm probe. The composition of metal ions Sr, Mn and Mo at different points of the grain was found to be $50\pm1$, $37\pm1$ and $13\pm1$ atomic percents respectively, which is close to the synthesized composition within the typical error of EDAX technique. Iodometric titration analysis revealed that the as prepared sample is oxygen stoichiometric. The value of off-stoichiometry $\delta$ was found to be only $-0.003\pm0.005$, which is well within our experimental error. Well characterized single phase samples were then subjected to structural phase-transition studies employing temperature dependent XRD ($83K$ to $637K$) and TEM ($100K$-$390K$). Thin sample for TEM studies was prepared using conventional method of ion-beam polishing[25]. For temperature dependent TEM studies GATAN make high-temperature holder 652 and liquid-nitrogen based low-temperature holder 636MA were used. To monitor the phase-transition four-probe electrical resistivity measurement was also carried out from $390K$ down to $90K$ during cooling and heating cycles both. For monitoring possible magnetic phase change magnetization-vs-temperature (*M-T*) measurement was carried out between $325K$-$600K$ using SQUID-VSM (QD).

**Results and Discussions:**

Fig.1 shows room temperature XRD profile of the as synthesized $SrMn_{0.875}Mo_{0.125}O_{3-\delta}$ and its insets show logarithmically scaled and enlarged views (a) of the XRD profile from $2\theta = 42°$ -$73°$ and (b) from $2\theta = 46°$-$48°$. Inset (a) shows occurrence of weak but clear diffraction peaks, as marked by arrows. The style of splitting of main peaks and their intensity ratios appear to indicate either a tetragonally or a pseudo tetragonally (i.e. orthorhombic) distorted perovskite lattice. A close inspection of shape and width of (h00)/(0k0) and (00l) profiles, e.g. (200)/(020) and (002) as shown in the inset (b), indicates that the combined width of (200)/(020) (= 0.155°) is slightly larger than that of the (002) (= 0.124°). This is due to the fact that the positions of the (200) and (020) basic perovskite reflections are slightly shifted and the basic perovskite is orthorhombically (pseudo tetragonally) distorted. Since the intensities of extra peaks are very weak, therefore, for a moment, ignoring the presence of these weak peaks, the XRD profile in Fig.1 is very nicely Rietveld refined ($\chi^2$=3.04) with an orthorhombic (pseudo tetragonal) phase with space-group Pmmm. The low $\chi^2$ value of the refinement accounting all the main peaks, indicates that the XRD profile belongs to a single phase sample of a pseudo tetragonally distorted perovskite phase with a=3.8671Å, b=3.8692Å and c=3.8259Å, superimposed with some kind of structural modulation. Nearly symmetrically placed weak peaks about the main perovskite reflections in a manganite sample has ingeneral been attributed to the CO superlattice modulation[14,25]. For confirming the CO modulation TEM investigations were carried out in imaging and diffraction modes. Figs. 2(a) & (b) show high-resolution lattice image and selected area electron diffraction (SAD) taken along [001] zone of the basic perovskite lattice of $SrMn_{0.875}Mo_{0.125}O_{3-\delta}$. Occurrence of strong superlattice spots along [-110] can be clearly seen. This is a characteristic of the occurrence of CO modulation. The fringes in Fig.2 (a) represent the direct image of the CO modulation. The A repeated careful measurement of the scattering vectors in the SAD pattern of Fig.2 (b) revealed that the $g_{-110}$ reciprocal lattice vector of the basic perovskite is a non-integral multiple of the modulation vector. The modulation wavelength is measured to be of 24.5 Å, which is about 8.95±0.01 times the $d_{-110}$. Thus the modulation vector **q** was estimated to be ~ 0.111. The current observation of CO modulation in $SrMn_{0.875}Mo_{0.125}O_{3-\delta}$ is broadly similar to the CO modulation observed in the analogous compound $CaMn_{1-x}Mo_xO_3$ for $x=0.12$ [46]. When compared in terms of CO modulation along [-110] of the basic perovskite, the modulation vector for the Ca based compound is also found to be nearly the same, q~ 0.1 to 0.125. The only difference in the two observations is the type of modification in the basic perovskite lattice. In our case the basic perovskite lattice remains the same but simply undergoes a pseudo tetragonal distortion, without involving any Glazer octahedral tilt[50]. While in the case of analogous Ca compound[46] the basic perovskite undergoes a Glazer octahedral tilt[50] leading to a Pnma structure. A careful observation of the micrograph in Fig.2(a) revealed the presence of waviness in the CO modulation fringes[31]. If we follow the fringe beneath the white line A in Fig.2(a), it goes straight. But the same is not true for the fringe beneath the line B, it slowly deviates down in the middle. Similarly if we follow the line C from left to right we find many fringes deviating from straight line in the region encircled as D. We also observe termination of a CO fringe in between two fringes in region encircled as E. This appears to represent an "edge dislocation" like situation in a CO modulation. Thus these microstructural observations present clear evidence of waviness of the CO modulation in the sample.

The replacement of 12.5% of $Mn^{+4}$ by $Mo^{+6}$ will create ~25% of $Mn^{+3}$ ions at the B-site. Thus the $Mn^{+4}$:$Mn^{+3}$ ratios become 3:1, which is expected to result a CO modulation wavelength of 8 times the $d_{-110}$. The occurrence of waviness and non-integral modulation indicate that the observed CO is an incommensurate CDW type modulation[28-31]. The Fig.2(c) shows [0-11] zone SAD pattern from the sample. Absence of diffraction spots at the center position (as indicated by arrow) of the rectangular arrangement of the basic perovskite reflections approves that there is no $GdFeO_3$ type structural distortion[50], which occurs due to tilting of $MO_6$ octahedra in a perovskite lattice giving rise to space groups modifications like e.g *I4/mcm, Imma, Pnma, P2$_1$/m,* etc. Absence of any such modification and the intactness of cubic symmetry is due to tolerance factor t ~ 1.03 arising due to the occupancy of A-site by large ionic size atom $Sr^{+2}$ (1.44Å). Thus currently the occurrence of charge and orbital ordering is the only phenomenon which is modifying the basic perovskite lattice. Fig.2 (d) shows a [001] zone SAD pattern taken at 100$K$. This shows that the CO phase is stable down to 100$K$. Fig.2 (e) represents intensity profile of a [-110] row showing perfect periodicity of the basic perovskite reflections. Unlike the case of 1-dimensional CO modulation observed earlier in $Pr_{1-x}Ca_xMnO_3$ and $La_{1-x}Ca_xMnO_3$ compounds, regions with 2-dimensional (2D) CO modulation, comprising mutually perpendicular modulation vectors, are also seen. High-resolution lattice image and the corresponding SAD pattern as shown in Fig.3 are the direct evidence of a 2D CO modulation. The occurrence of 2D CO (CDW) has not been reported earlier. This appears to reflect upon some characteristic difference in the nature of orbital ordering in this compound. Fig.4 shows resistivity-vs-temperature (R-T) variation of $SrMn_{1-x}Mo_xO_3$ ($x$=0.125). This is similar to the $\rho$-$T$ data shown by Maignan et al.[47]. The step like variation in $R$-$T$ represents CO induced MI transition at 355$K$. Occurrence of a broad hysterisis ranging from 150$K$ to 355$K$ in the $R$-$T$ can be clearly seen. This indicates that the observed CO transition is accompanied with a broad first-order structural phase-transition. The inset of Fig.4 shows magnetization-vs-temperature ($M$-$T$) variation from 325$K$ to 600$K$. The sharp drop in $M$-$T$ at ~350$K$ corresponds to AFM transition. Fig.5 shows [-110] row of the [001] zone SAD patterns recorded at different temperatures during heating. It can be seen that above 354$K$ the intensity and sharpness of the CO superlattice reflections start diminishing quite drastically and almost vanishes at ~375$K$ leaving behind its reminiscence as diffuse cross strikes. Thus the observed superlattice spots are fully correlated with the MI transition in $R$-$T$ and AFM transition in $M$-$T$ and therefore ascertain that they belong to charge ordering.

In manganites the occurrence of CO has been attributed to both, longitudinal and transverse modulation of atomic positions with respect to the modulation wave vector[14,21]. To probe the type of modulation in the present case we followed the technique as described by Nagai et al[51] and collected off-tilted [001] zone SAD pattern by tilting the sample about [-110] direction, see the SAD in Fig.6. It can be seen that all the CO modulation spots with modulation vector q=0.1117 along [-110] have almost vanished, indicating that they were kinematically forbidden and appeared merely due to double diffraction. The kinematical absence of the CO spots directly implies that the corresponding atomic position modulation is transverse to the CO modulation vector. This implies that the observed CO is of "Wigner crystal" type as reported by Radali et al[14], Wang et al[21] and Nagai et al[51].

To confirm the bulk occurrence of the observed CO phase and also to measure the related structural changes more accurately, temperature dependent powder XRD was carried out at various temperatures between 83$K$ and 700$K$. The obtained XRD data were Rietveld refined. Fig.7 shows some selected XRD profiles at 682$K$, 295$K$ and 83$K$. The XRD profile at 682K as shown in Fig.7(a) was refined with cubic phase with space-group *Pm-3m*. A comparison of the degree of resolution of $K_{\alpha 1}$ and $K_{\alpha 2}$ lines of (400)/(040) &(004) reflections of the XRD profiles at 295$K$ and 83$K$, as shown in the insets of Fig.7(b)&(c), clearly indicates a pseudo tetragonal (or orthorhombic) type distortion of the basic perovskite. Thus the XRD profiles at 295$K$ and 83$K$, as shown in Fig.7(b) & (c), were best refined with an orthorhombic phase with space-group *Pmmm*. Fig.8 exhibits the plot of temperature variation of the lattice parameters of the *Pm-3m* and *Pmmm* phases across the phase-transition. The cubic perovskite transforms to orthorhombically (pseudo tetragonal) distorted lattice at 355$K$. It can be noticed that across CO transition below 355$K$ the *a* & *b* lattice parameters increase while the *c* parameter decreases. The increase of *a/b* parameters almost saturates at ~275$K$ whereas the *c* parameter keeps decreasing till 100$K$. The observed anisotropy in the thermal variation of *a/b* and *c* lattice parameters below 255$K$ appears to arise due to combined effect of lattice changes due to phase transition; i.e. due to orbital-ordering, and that of due to typical thermal effects on the basic perovskite structure, as typified by the linear variation of the lattice parameter in its cubic phase regime above 360$K$. The increasing anisotropy in the lattice parameter down to 150$K$ thus indicates a broad phase transition. This is in accordance with the large hysterisis observed in the *R-T* data as shown in Fig.4. Fig.9 shows stack of slow scanned XRD profiles about (121/211) and (112) reflections (under *Pmmm* consideration) taken at close temperature intervals across the phase transition. Melting of the CO phase in the bulk sample, as indicated by arrows, can be seen. Unlike electron diffraction observation, the CO superlattice peaks in XRD appear to vanish at 332*K*. This is due to very weak intensity of the superlattice reflections above 332*K*. But the true indicator of the phase transitions, the splitting of (121/211) and (112) reflections, vanishes only above 355*K*. This approves that the commencement of charge ordering and the occurrence of pseudo tetragonal distortion of the cubic lattice are intimately related.

The above described results clearly show that incommensurate CO transition occurs in $SrMn_{0.875}Mo_{0.125}O_3$ above room temperature at 355K resulting in an antiferromagnetic insulating phase. X-ray diffraction results approve that the observed CO phase is a bulk occurrence and it results in a coupled structural transition from cubic to pseudo tetragonal due to associated orbital ordering. The accurate accounting of all the main reflections in the Rietveld refinement of the low temperature XRD data, by a simple perovskite lattice, as well as the perfect periodicity of the perovskite reflections, as evident from SADs and corresponding intensity profile in Fig.2 (e), directly imply that the cubic to pseudo tetragonal transition is a consequence of $c/a < 1$ type coherent distortion occurring in each unit cell. Such a distortion of the perovskite lattice at first site appears to suggest a $d_{x^2-y^2}$ type orbital ordering confined in the a-b plane. But the simultaneous occurrence of CO modulation suggests another possibility as well, i.e. of herringbone type ordering of $3d_x^2/3d_y^2$ orbitals in the *a-b* plane, which is usually seen in the case of CE-type CO antiferromagnetic phase. Schematically these models are presented in Fig.10 (a) and (b). Both these orbital ordering models, since the corresponding $e_g$ charge densities are confined in the *a-b* plane, will cause expansion of the a-b plane. Therefore, under the constrain that the unit cell volume should remain constant; the *c*-axis gets shrunk. Lack of $e_g$ charge density along *c*-axis further facilitates this shrinkage. That is what has been experimentally observed.

The CO phase modeled as herringbone type ordering of $3d_x^2/3d_y^2$ orbitals always results in a 1D modulation propagating along [-110] of the basic perovskite, as observed in the high-resolution electron micrograph in Fig.2(a). A planer ordering of $3d_{x^2-y^2}$ orbitals usually results in a ferromagnetic interaction in the *a-b* plane, resulting in A-type AFM phase. But Khomskii et al [52] and Mutou et al[53] have theoretically shown that occurrence of long-range coulomb interaction in half doped manganites does result a COO ground state, which is based on planer ordering of $3d_{x^2-y^2}$ type orbitals confined in the a-b plane, as shown in Fig.10(a). Therefore, in the present case, where the basic perovskite lattice undergoes a pseudo tetragonal distortion of *c/a<1* type, the case of 1D CO modulation with $3d_{x^2-y^2}$ type orbital ordering becomes a logical possibility[52,53]. Here it appears imperative to point out that the basic perovskite lattice of most of the CE-type charge ordered manganites with $d_x^2/d_y^2$ type orbital ordering, involves octahedral tilt[50] which is essential to minimize the unit cell volume and to accommodate the orbital arrangement. The absence of any such octahedral tilt in our samples is likely to favor $3d_{x^2-y^2}$ type orbital ordering. Further the experimental observation of 2D CO modulation, as shown in Fig.3 (a), which demands transverse atomic modulation in two mutually perpendicular directions, makes this possibility still stronger.

Fig.10(c) & (d) show models for 8-times modulated 2D COO phases with $3d_{x^2-y^2}$ and $3d_x^2/3d_y^2$ type orbitals respectively. In these models the COO modulation vectors are shown going horizontally and vertically both. For the model given in Fig.10 (d) the

$3d_x^2/3d_y^2$ orbital strips, which are aligned vertically and are associated with the horizontal 8-fold COO modulation, have normal herring bone type arrangement. But the $3d_x^2/3d_y^2$ orbital strips, which are aligned horizontally and are associated with the vertical 8-fold COO modulation, have mismatched herring bone arrangement of $3d_x^2/3d_y^2$ orbitals at its interface with the vertical orbital strips. This "mismatching" of $3d_x^2/3d_y^2$ orbitals is like defect in the ordering and is present in each unit cell. Few of these defects are encircled in Fig.10 (d). Moreover this mismatch of the orbitals will cause interruption in the delocalization of $e_g$ charge densities along the zig-zag chain, which in turn will hinder the ferromagnetic ordering of the spins along the zig-zag chain. The presence of these defects are likely to cause larger energy cost for the stabilization of a mutually perpendicular herring bone type arrangement of $3d_x^2/3d_y^2$ orbitals and therefore a 2D COO modulation involving $3d_x^2/3d_y^2$ type orbitals is unlikely. On the contrary, the $3d_{x^2-y^2}$ orbitals, due to its 4-fold symmetric $e_g$ charge density, will neither cast any such orbital misfit defect at the interfaces of the mutually perpendicular crossing modulation vectors nor will cause any hindrance to the ferromagnetic spin ordering in zig-zag chains. This is very much clear from the model given in Fig. 10 (c). Based on these discussions we argue that $3d_{x^2-y^2}$ orbitals are the natural selection for the occurrence of a 2D COO modulation than the $3d_x^2/3d_y^2$ type orbitals. The mutually perpendicular and identical charge distribution lobes of $3d_{x^2-y^2}$ orbitals will facilitate/support mutually perpendicular modulation resulting in 2D COO. The present case of coexisting 1D and 2D COO phases appears to be a case of coexistence of a CO phase with usual CE type spin ordering associated with $3d_x^2/3d_y^2$ type orbital ordering and a CO phase associated with $3d_{x^2-y^2}$ type 2D orbital-ordering.

**Conclusion:**

Based on the above results and discussions it is concluded that B-site Mo doped disordered perovskite $SrMn_{0.875}Mo_{0.125}O_3$ shows a strong charge-ordered antiferromagnetic insulating phase at room temperature in its pseudo tetragonally distorted perovskite phase. This charge ordering appears to be an incommensurate charger density wave type transition causing transverse atomic modulation of the basic cubic perovskite phase at ~355$K$. The occurrence of pseudo tetragonality with c/a<1 and the regions with 2D CO modulation, as seen through TEM, has been analyzed as 2D CO modulation with $3d_{x^2-y^2}$ type orbital ordering in the a-b plane. The coexistence of regions with 1D and 2D CO modulations reveals the coexistence of herring bone type arrangement of $3d_x^2/3d_y^2$ type orbitals and planer ordering of $3d_{x^2-y^2}$ type orbitals in the basic perovskite structure.

**Acknowledgment**

Authors gratefully acknowledge Dr. A. K. Sinha, Dr. V. Ganesan, Dr. P. Chaddah and Prof. A. Gupta for their constant encouragement and support. Dr. R. J. Choudhary is acknowledged for magnetization measurements. Aga Shahee would like to acknowledge CSIR-India for financial support as SRF.

**References:**

[1] J. G. Bednorz and K. A. Müller, Z. Physik, B **64,** 189–193(1986).

[2] D. J. Van Harlingen. Rev. Mod. Phys. **67**, 515 (1995)

[3] S. Jin, T. H. Tiefel, M. McCormack, R. A. Fastnacht, R. Ramesh, and L. H. Chen, Science **264**, 413 (1994).

[4] K. H. Ahn, T. Lookman, and A. R. Bishop, Nature (London) **428**, 401 (2004).

[5] J. B. Goodenough, Phys. Rev. **100,** 564 (1955).

[6] R. Maezono, S. Ishihara, and N. Nagaosa, Phys. Rev. B **58**, 11583 (1998).

[7] R. Maezono, S. Ishihara, and N. Nagaosa, Phys. Rev. B **57**, R13993 (1998).

[8] D. I. Khomskii, and K. I. Kugel, Phys. Rev. B **67**, 134401 (2003).

[9] A P Ramirez, J. Phys.: Condens. Matter **9**, 8171–8199(1997).

[10] E. Dagotto, T. Hotta, and A. Moreo, Phys. Rep. **344**, 1 (2001).

[11] P. Schiffer, A. P. Ramirez, W. Bao, and S-W. Cheong, Phys. Rev. **75**, 3336 (1995).

[12] M. Uehara, S. Mori, C. H. Chen, and S. W. Cheong, Nature (London) **399**, 560 (1999).

[13] C. H. Chen and S-W. Cheong Phys. Rev. Lett. **76**, 4042 (1996).

[14] P. G. Radaelli, D. E. Cox, M. Marezio, and S.-W. Cheong, Phys. Rev. B **55**, 3015 (1997).

[15] C. N. R. Rao and A. K. Cheetham, Science **276**, 911 (1997).

[16] S. Mori, C. H. Chen, and S.-W. Cheong, Nature (London) **392**, 473 (1998).

[17] S. Mori, C. H. Chen and S-W. Cheong, Phys. Rev. Lett. **81**, 3972 (1998).

18. G. Allodi, R. De Renzi, F. Licci and M. W. Pieper, Phys. Rev. Lett. **81**, 4736 (1998).
19. P. G. Radaelli, D. E. Cox, L. Capogna, S.-W. Cheong and M. Marezio Phys. Rev. B **59**, 14440 (1999).
20. S. Mori, T. Katsufuji, N. Yamamoto, C. H. Chen and S-W. Cheong, Phys. Rev. B **59**, 13573 (1999).
21. R. Wang, J. Gui, Y. Zhu and A. R. Moodenbaugh, Phys. Rev. B **61**, 11946 (2000).
22. C. N. R. Rao, J. Phys. Chem. B **104**, 5877 (2000).
23. R. Kajimoto, H. Yoshizawa, Y. Tomioka and Y. Tokura, Phys. Rev. B **63**, 212407 (2001).
24. X. G. Li1, R. K .Zheng, G. Li ,H. D. Zhou, R. X. Huang, J. Q. Xie and Z. D. Wang, Europhys. Lett., **60**, 670 (2002).
25. P. R. Sagdeo, Shahid Anwar, and N. P. Lalla. Phys. Rev. B **74**, 214118 (2006).
26. L. Wu, R. F. Klie, Y. Zhu, and Ch. Jooss, Phys. Rev. B **76**, 174210 (2007).
27. P. R. Sagdeo, N. P. Lalla, A. V. Narlikar, D. Prabhakaran, and A. T. Boothroyd Phys. Rev. B **78**, 174106 (2008).
28. J. R. Carvajal, A. D. Aladine, L. P. Gaudart, M. T. F. Díaz and A. Revcolevschi, Physica B **320**, 1 (2002).
29. J. C. Loudon, S. Cox, A. J. Williams, J. P. Attfield, P. B. Littlewood, P. A. Midgley, and N. D. Mathur, Phys. Rev. Lett. **94**, 097202 (2005).
30. G. C. Milward, M. J. Calder'on, and P. B. Littlewood, Nature, **433**, 607 (2005).

31   D. Sánchez, J. Calderón, J. Sánchez-Benítez, A. J. Williams, J. P. Attfield, P. A. Midgley and N. D. Mathur, Phys. Rev. B, 77, 092411 (2008).

32   M. Hervieu, A. Maignan, C. Martin, N. Nguyen, and B. Raveau, Chem. Mater., **13,**1356 (2001).

33   Ch. Renner, G. Aeppli, B.-G. Kim, Yeong-Ah Soh and S.-W. Cheong, Nature, 416, 518 (2002).

34   M. Giot, P. Beran, O. Peŕez, S. Malo, M. Hervieu, B. Raveau, M. Nevriva, K. Knizek, and P. Roussel, Chem. Mater. **18**, 3225 (2006).

35   J. L. García-Muñoz, C. Frontera, B. Rivas-Murias, and J. Mira, J. Appl. Phys. **105**, 084116 (2009).

36   P. W. Anderson, Phys. Rev. **109**, 1492 (1958).

37   J. M. D. Coey, M. Viret, and S. Von Molnár, Adv. Phys. **48**, 167 (1999).

38   E. Dagotto, T. Hotta, and A. Moreo, Phys. Reports **344**, 1 (2001).

39   J. Lee, B. Kim, B. H. Kim, B. I. Min, S. Kolesnik, O. Chmaissem, J. Mais, B. Dabrowski, H. J. Shin, D. H. Kim, H. J. Lee, and J.-S. Kang, Phys. Rev. B **80**, 205112 (2009).

40   A. Maignan, C. Martin, M. Hervieu, and B. Raveau, J. Appl. Phys. **91**, 4267 (2002).

41   A. Barnabé, A. Maignan, M. Hervieu, F. Damay, C. Martin and B. Raveau, Appl. Phys. Lett. **71**, 3907 (1997).

42   B. Raveau, A. Maignan, and C. Martin, J. Solid State Chem. **130**, 162 (1997).

43   T. Kimura, Y. Tomioka, R. Kumai, Y. Okimoto, and Y. Tokura, Phys. Rev. Lett. **83**, 3940 (1999).

44  C. Yaicle, C. Martin, Z. Jirak, F. Fauth, G. André, E. Suard, A. Maignan, V. Hardy, R. Retoux, M. Hervieu, S. Hébert, B. Raveau, Ch. Simon, D. Saurel, A. Brûlet, and F. Bourée, Phys, Rev. B **68**, 224412 (2003).

45  A. Banerjee, K. Mukherjee, Kranti Kumar, and P. Chaddah, Phys. Rev. B **74**, 224445 (2006).

46  C. Martin, A. Maignan, M. Hervieu, B. Raveau, and J. Hejtmanek, Phys. Rev. B **63**, 100406R (2001).

47  A. Maignan, C. Martin, M. Hervieu, and B. Raveau, J. Appl. Phys. **91**, 4267 (2002).

48  T. Suzuki, H. Sakai, Y. Taguchi, and Y. Tokura, J. Electronic Mat. **41**, 1559 (2012).

49  J. Lee, B. Kim, B. H. Kim, B. I. Min, S. Kolesnik, O. Chmaissem, J. Mais, B. Dabrowski, H. J. Shin, D. H. Kim, H. J. Lee, and J.-S. Kang, Phys. Rev. B **80**, 205112 (2009).

50  A. M. Glazer, Acta Cryst. B **28**, 3384 (1972).

51  T. Nagai, T. Kimura, T. Asaka, K. Kimoto, M. Takeguchi, and Y. Matsui, Phys. Rev. B **81**, 060407(R) (2010).

52  D. Khomskii and J. van den Brink, Phys. Rev. Lett. **85,** 3329 (2000).

53  T. Mutou, and H. Kontani, Phys. Rev. Lett. **83**, 3685 (1999).

**Figure caption:**

**Figure 1**. Rietveld refined room temperature XRD data of $SrMn_{0.875}Mo_{0.125}O_3$ with *Pmmm* space group. Inset (a) shows logarithmically scaled region of the room temperature XRD data from $42^o$-$73^o$ $2\theta$. Presence of weak superlattice spots can be clearly seen. Inset (b) shows enlarged view of (200)/(020) and (002) peaks. Comparatively ill resolved $K\alpha_1$ & $K\alpha_2$ lines for (200)/ (020) reflections than that of the (002) peak indicate pseudo tetragonal distortion of the basic perovskite lattice.

**Figure 2.** (a) High resolution electron micrograph and (b) corresponding [001] zone SAD pattern of charge-ordered (CO) superlattice modulation observed at room temperature in $SrMn_{0.875}Mo_{0.125}O_3$. (c) [0-11] zone SAD pattern showing absence of diffraction spots essentially observed in the case of $GdFeO_3$ type perovskite distortion.(d) [001] zone SAD pattern taken at 100K. This exhibits the stability of the observed CO phase at lower temperatures. (e) Intensity profile along [-110] row of the [001] zone SAD pattern showing CO peaks and periodically spaced basic perovskite peaks.

**Figure 3.** (a) High resolution electron micrograph and (b) corresponding zone axis SAD pattern showing occurrence of 2D modulated charge-ordered regions in $SrMn_{0.875}Mo_{0.125}O_3$ .

**Figure 4.** Temperature variation of resistance of $SrMn_{0.875}Mo_{0.125}O_3$ between 85*K* to 385*K* showing metal-insulator (*MI*) transition at 355*K* with a broad hysterisis ranging from ~150*K* to 365*K*. Inset shows the *M-T* data showing AFM transition accompanying the MI transition.

**Figure 5.** Stack of [-110] row SAD patterns of $SrMn_{0.875}Mo_{0.125}O_3$ taken during heating from 295*K* -376*K*. CO superlattice spots disappear at 354*K*.

**Figure 6.** Off tilted [001] zone SAD pattern taken after tilting the sample about [-110] direction. Absence of the CO superlattice spots along [-110] proves their kinematical absence.

**Figure 7.** Rietveld refined XRD data of SrMn$_{0.875}$Mo$_{0.125}$O$_3$ taken at (a) 682$K$ (*Pm-3m*) (b) 295K (*Pmmm*) and (c) 83K (*Pmmm*). Insets show enlarged view (400) type reflections. The inset in (c) very clearly shows comparatively ill resolved K$\alpha_1$ & K$\alpha_2$ lines in (400)/ (040) reflections than that of the (004).

**Figure 8.** Temperature variation of the basic perovskite lattice across the cubic (*Pm-3m*) to Pmmm phase transition. Highly anisotropic nature of thermal dependence of the a/b and c parameters can be observed.

**Figure 9.** Slow scanned XRD profiles of SrMn$_{0.875}$Mo$_{0.125}$O$_3$ around (121/211) and (112) reflections taken during heating from 83$K$ to 367$K$. Melting of CO superlattice peaks above 354$K$ can be seen.

**Figure 10.** Models showing possible (a) 1D and (d) 2D charge orbital ordered (COO) states comprising 3$d_{x2-y2}$ type orbitals arising due to long range coulomb interaction and (b) 1D COO model, the usual one with CE type magnetic structure and (c) 2D COO model with ordering of 3$d_{x2}$/3$d_{y2}$ type orbitals.

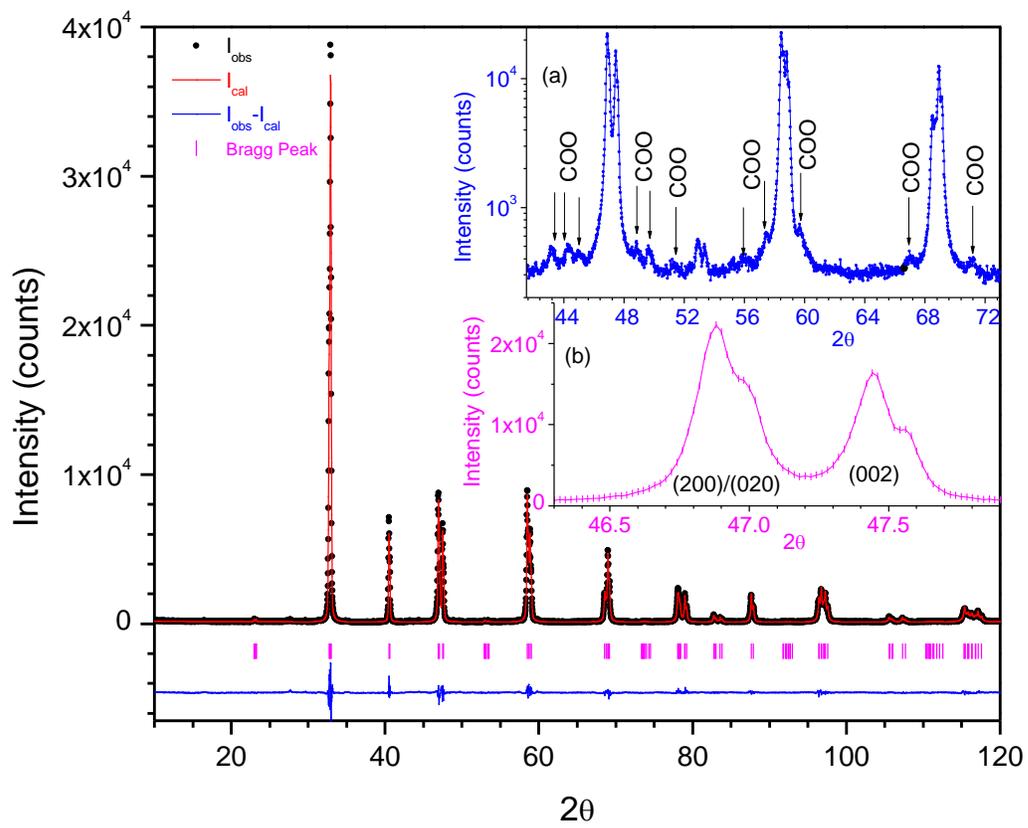

Figure 1

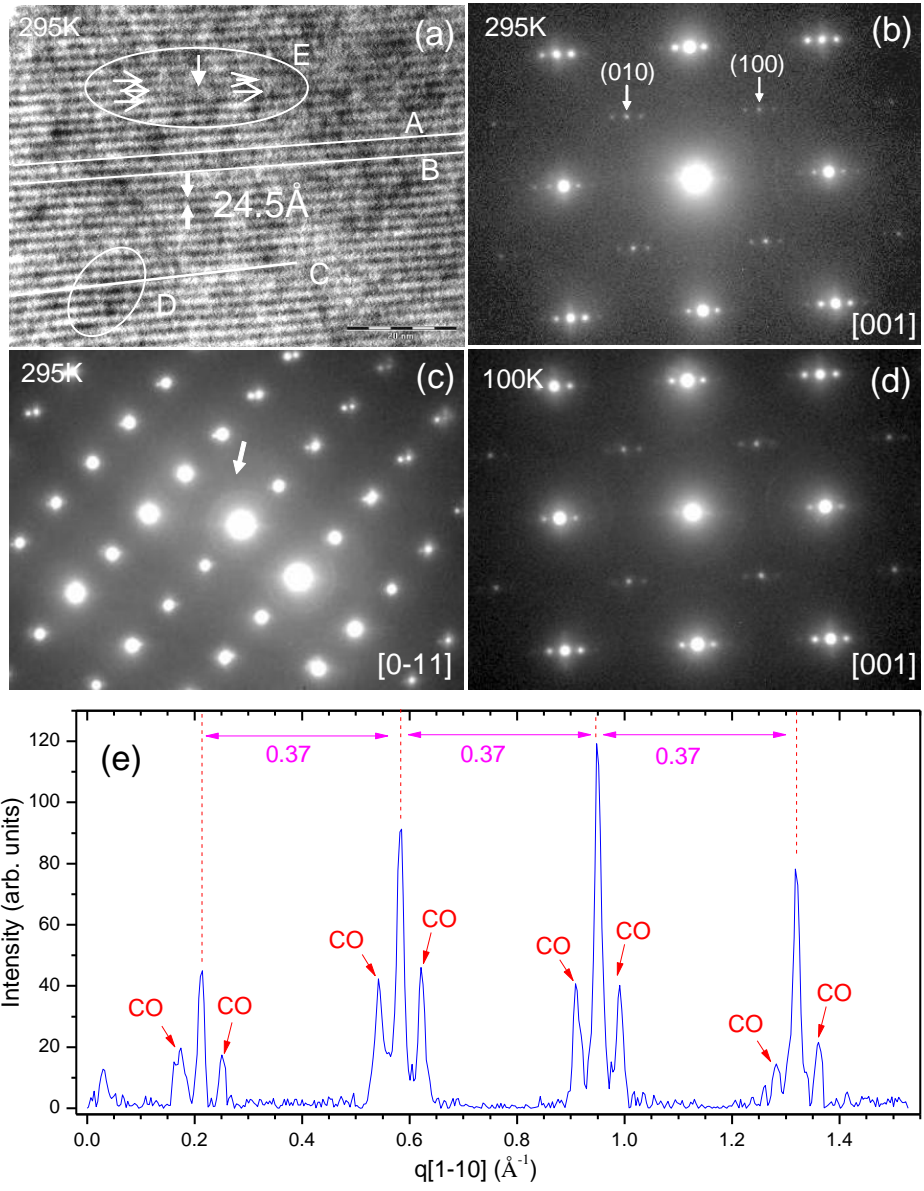

Figure 2

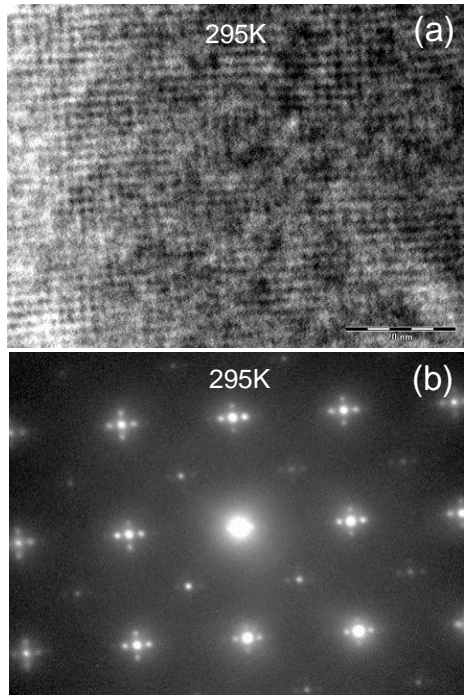

Figure 3

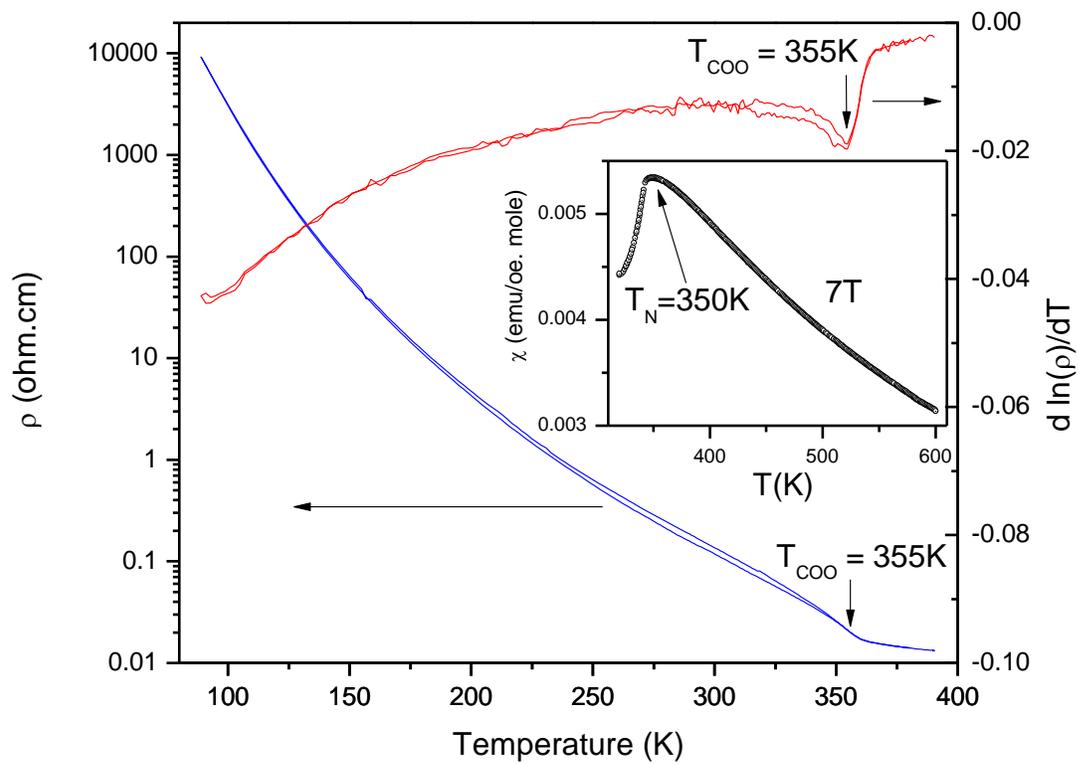

Figure 4

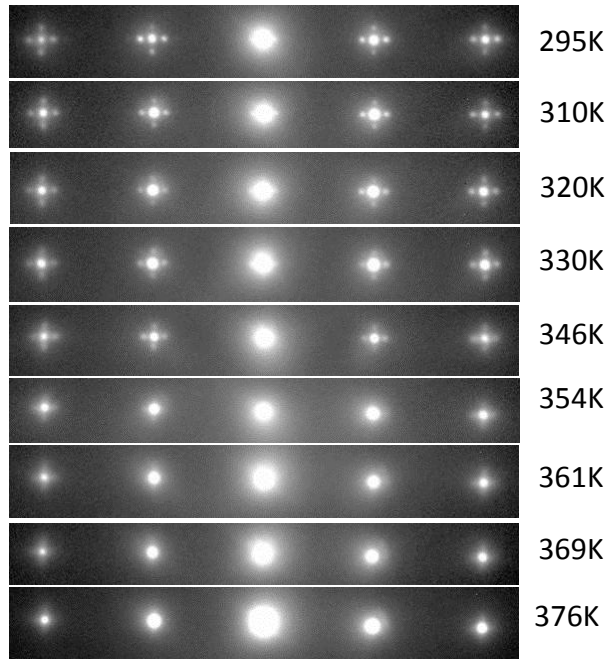

Figure 5

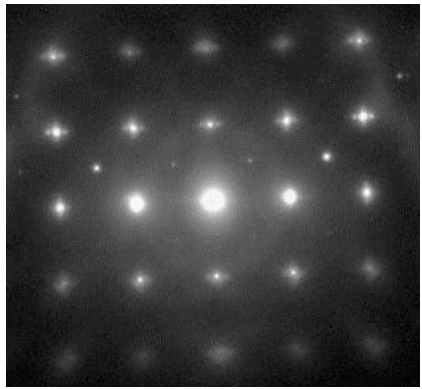

Figure 6

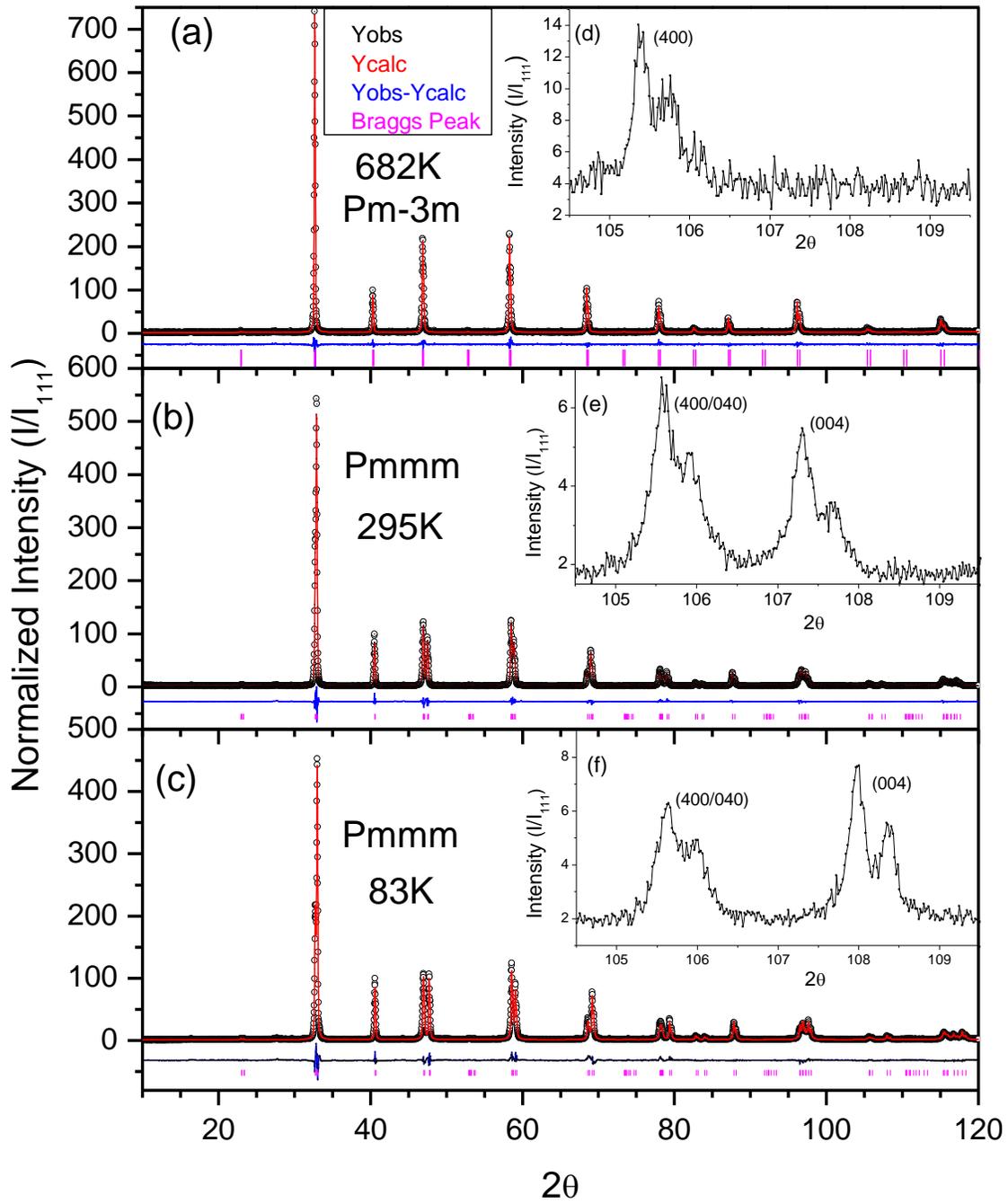

Figure 7

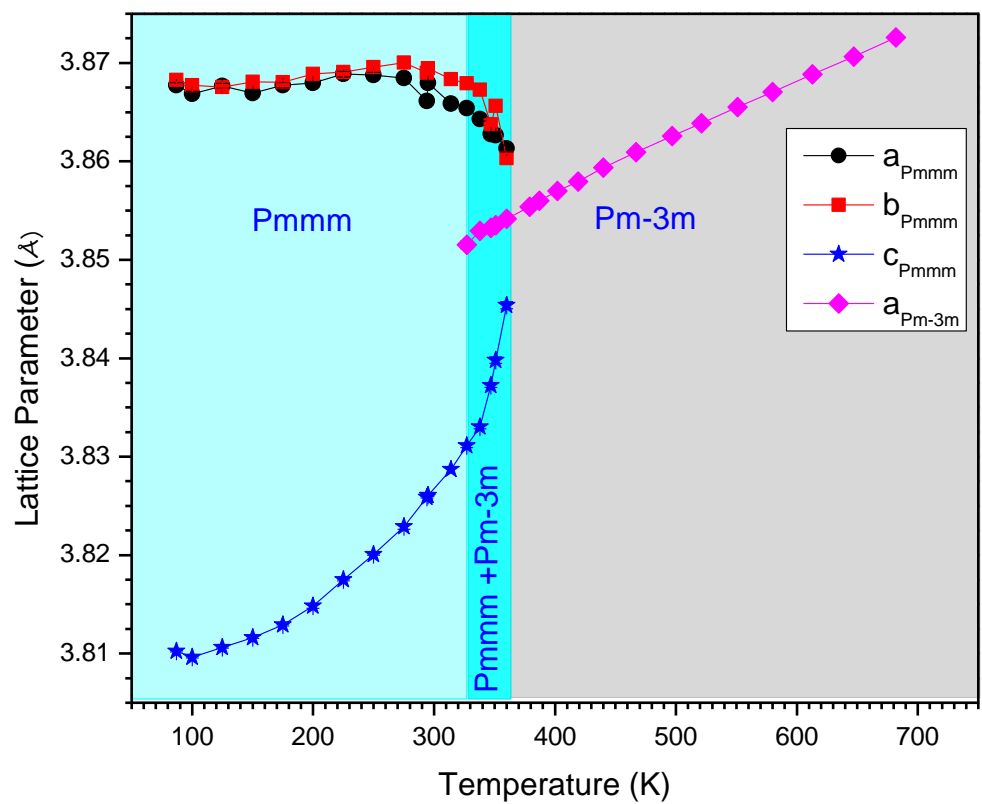

Figure 8

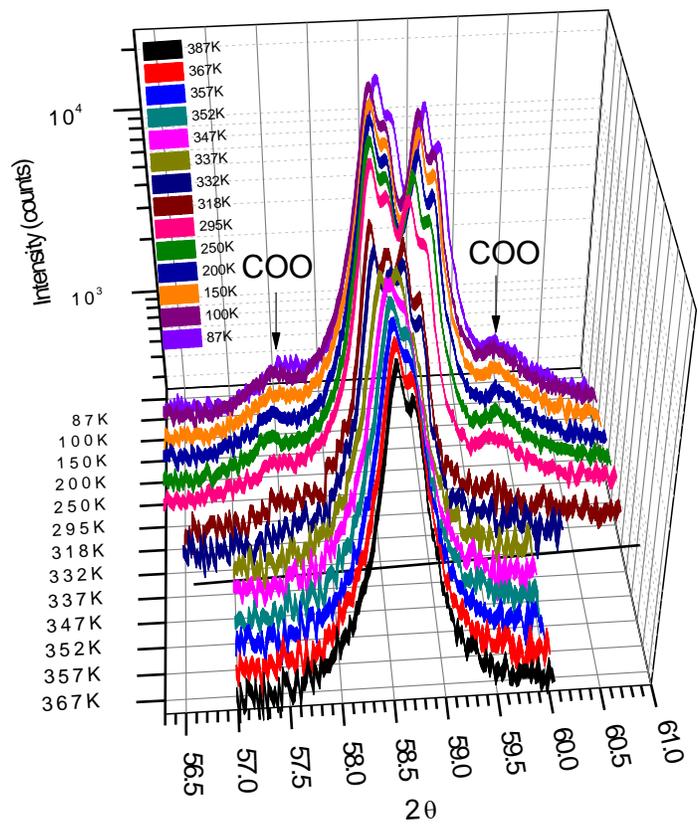

Figure 9

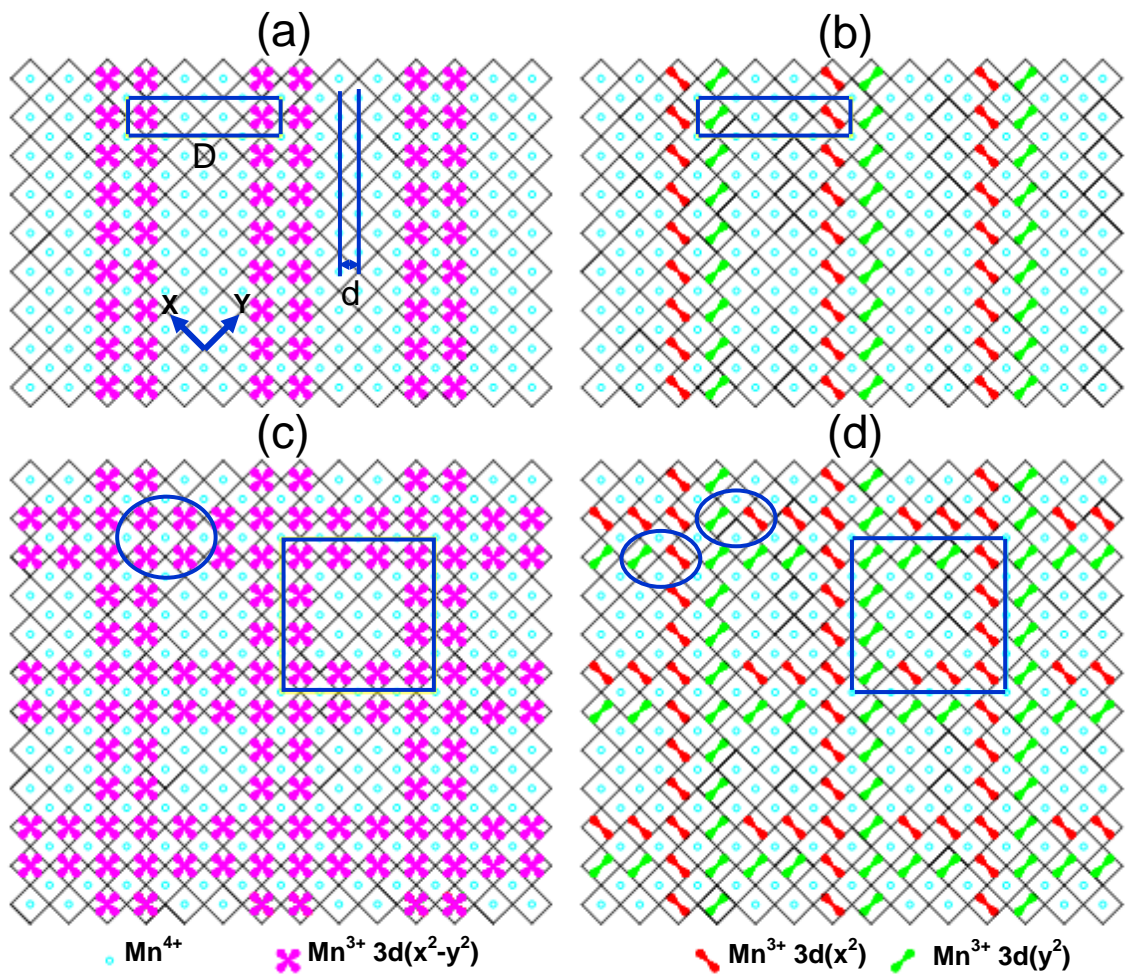

Figure 10